\documentclass{jfm}

\usepackage{graphicx}
\usepackage{natbib}
        
\ifCUPmtlplainloaded \else
  \checkfont{eurm10}
  \iffontfound
    \IfFileExists{upmath.sty}
      {\typeout{^^JFound AMS Euler Roman fonts on the system,
                   using the 'upmath' package.^^J}%
       \usepackage{upmath}}
      {\typeout{^^JFound AMS Euler Roman fonts on the system, but you
                   dont seem to have the}%
       \typeout{'upmath' package installed. JFM.cls can take advantage
                 of these fonts,^^Jif you use 'upmath' package.^^J}%
      }
  \else
  \fi
\fi

\ifCUPmtlplainloaded \else
  \checkfont{msam10}
  \iffontfound
    \IfFileExists{amssymb.sty}
      {\typeout{^^JFound AMS Symbol fonts on the system, using the
                'amssymb' package.^^J}%
       \usepackage{amssymb}%
         \let\leq=\leqslant
         \let\geq=\geqslant
      }{}
  \fi
\fi

\ifCUPmtlplainloaded \else
  \IfFileExists{amsbsy.sty}
    {\typeout{^^JFound the 'amsbsy' package on the system, using it.^^J}%
     \usepackage{amsbsy}}
    {}
\fi

\newsavebox{\astrutbox}
\sbox{\astrutbox}{\rule[-5pt]{0pt}{20pt}}

\usepackage{amsmath}
\usepackage{multirow}

\title{Measurements of the coupling between the tumbling of rods and the velocity gradient tensor in turbulence}

\author[R. Ni, S. Kramel, N. T. Ouellette and G. A. Voth]
{R\ls U\ls I\ns  N\ls I\ns$^{1,2}$\thanks{Email address for correspondence: ruiniphy@gmail.com,}, S\ls T\ls E\ls F\ls A\ls N\ns  K\ls R\ls A\ls M\ls E\ls L\ns$^1$, N\ls I\ls C\ls H\ls O\ls L\ls A\ls S\ns T. \ls O\ls U\ls E\ls L\ls L\ls E\ls T\ls T\ls E$^{2}$ \and G\ls R\ls E\ls G\ns \ls A. V\ls O\ls T\ls H$^1$%
 \thanks{Email address for correspondence: gvoth@wesleyan.edu}}

\affiliation{$^1$Department of Physics, Wesleyan University, Middletown, Connecticut 06459, USA
\\[\affilskip]
$^2$Department of Mechanical Engineering \& Materials Science, Yale University, New Haven, Connecticut 06520, USA}

\begin{document}
\maketitle


\begin{abstract}

We present simultaneous experimental measurements of the dynamics of anisotropic particles transported by a turbulent flow and the velocity gradient tensor of the flow surrounding them. We track both rod-shaped particles and small spherical flow tracers using stereoscopic particle tracking.   By using scanned illumination, we are able to obtain a high enough seeding density of tracers to measure the full velocity gradient tensor near the rod. The alignment of rods with the vorticity and the eigenvectors of the strain rate show agreement with numerical simulations. A full description of the tumbling of rods in turbulence requires specifying a seven-dimensional joint probability density function (PDF) of five scalars characterizing the velocity gradient tensor and two scalars describing the relative orientation of the rod. If these seven parameters are known, then Jeffery's equation specifies the rod tumbling rate and any statistic of rod rotations can be obtained as a weighted average over the joint PDF. To look for a lower-dimensional projection to simplify the problem, we explore conditional averages of the mean-squared tumbling rate. The conditional dependence of the mean-squared tumbling rate on the magnitude of both the vorticity and the strain rate is strong, as expected, and similar. There is also a strong dependence on the orientation between the rod and the vorticity, since a rod aligned with the vorticity vector tumbles due to strain but not vorticity. When conditioned on the alignment of the rod with the eigenvectors of the strain rate, the largest tumbling rate is obtained when the rod is oriented at a certain angle to the eigenvector that corresponds to the smallest eigenvalue, because this particular orientation  maximizes the contribution from both the vorticity and strain.
\end{abstract}

\section{Introduction}

Particles carried by turbulent flows in nature, such as ice crystals in clouds \citep{2003JASKorolev,2003ShawRA,2007Pinsky} or plankton in the oceans \citep{2007LimnolBoss, 2009Boss, 2012ARFMJeff}, are rarely spherical; instead, they often have nontrivial, anisotropic shapes that may influence their motion as they are carried by the flow. The simplest shape to consider after a sphere is that of an ellipsoid. The tumbling rate of an axi-symmetric ellipsoid in Stokes flow is determined by the particle orientation and the velocity gradient tensor~\citep{1922PRSLAJeffery}:   
\begin{equation}
\dot{p}_i=\Omega_{ij}p_j+\frac{\alpha^2-1}{\alpha^2+1}(S_{ij}p_j-p_ip_kS_{kl}p_l),
\label{eq:Jeff}
\end{equation}
where ${\bf{\hat{p}}}$ is a unit director along the symmetric axis of the particle, $\alpha$ is the aspect ratio of the ellipsoid, and $S_{ij}$ and $\Omega_{ij}$ are the symmetric and antisymmetric parts of the velocity gradient tensor, respectively.   

In a turbulent flow, a small ellipsoidal particle rotates in response to the velocity gradients along its Lagrangian trajectory.  Because these Lagrangian velocity gradients are controlled by the small scales, they are similar in many different turbulent flows and have been the focus of extensive study~\citep{2011ARFMMeneveau}.    To understand the dynamics of ellipsoidal particles in turbulence, we need to extend our understanding of the Lagrangian statistics of the velocity gradient tensor to include the orientational dynamics that result from integrating equation~\ref{eq:Jeff} along particle trajectories.   This is a challenging problem, both because of the complexity of statistically quantifying the particle orientation with respect to the velocity gradient tensor, and due to the difficulty of measuring the dynamics of anisotropic particles simultaneously with the velocity gradient tensors.  

The study of anisotropic particles in fluid flows has a long history because of the many relevant applications.  \citet{1980Leal} provides a review of the older literature, and a wide range of work has followed, for example: ~\citet{Koch1989, Szeri1993,Herzhaft1996, Olson1998,Parsa2011,Rosen2014, Andersson2013}.  Turbulent flows advecting anisotropic particles provide a compelling test case, both because of the many applications and because of the nearly universal statistics of the velocity gradients experienced by small particles in many turbulent flows at large Reynolds number.  However, the difficulty of accessing particle and fluid variables in turbulent flows has hindered work in this area.   Rod-shaped particles ($\alpha \gg 1$) were the first to be studied.  \citet{2005Shin} provided an extensive numerical study of rotational diffusion and the tumbling rate of rods in turbulence.  They observed that the tumbling rate of rods is much smaller than that predicted for randomly oriented rods.  \citet{2011NJPPumir} showed from numerical simulations that this suppression of the tumbling rate is caused by rods aligning with the vorticity vector.  \citet{2012PRLShima} extended numerical study of the tumbling rate across the full range of aspect ratios of axi-symmetric ellipsoids and found that preferential alignment decreases the tumbling rate for almost all shapes.  They also provided the first time-resolved experimental measurements of tumbling of rods in turbulence.    \citet{2013JFMMeneveau} studied the full parameter space of tri-axial ellipsoids in numerical simulations and showed that the tumbling of rods is a challenging test case for stochastic models of the velocity gradient tensor in turbulence.  \citet{2014PRLMehlig} used analytical and numerical methods to show the differences in tumbling between rods and disks can be understood using Lagrangian three-point correlations of the velocity gradient tensor.    \citet{2014PRLShima} made experimental measurements of the rotation of rods with lengths extending into the inertial range of turbulence and proposed that rotations of long rods should show inertial range scaling. \citet{2014JFMNi} showed how Lagrangian stretching aligns the long axis of a particle with the vorticity in turbulence.  

A question that remains unanswered is how to quantify the preferential orientation of particles that decreases their tumbling rate in turbulence.  From numerical simulations~\citep{2011NJPPumir,2014PRLMehlig} we know the probability distributions of the projection of $\bf{\hat{p}}$ onto the vorticity and the eigenvectors of the strain rate.  But predicting the tumbling rate requires the full joint probability distribution of the velocity gradient tensor and the particle orientation. This could be obtained from numerical simulations, although a thorough study has not yet been published.  In this paper we show that this joint probability distribution is now accessible to experimental measurements. 

One of the main challenges in simultaneous measurement of the orientations of particles and the velocity gradient tensor along the particle trajectory is that measuring  the Lagrangian velocity gradient tensor is difficult and typically has large experimental uncertainties.  This is primarily because the Kolmogorov length scale $\eta$ over which the flow is roughly linear is small, typically tens or hundreds of microns at high Reynolds numbers. Achieving this spatial resolution in an experiment is highly nontrivial. Multi-sensor hot-wire probes \citep{2010ARFMWallace} can provide the required resolution, but acquire only single-point Eulerian information and typically require the use of Taylor's hypothesis and a strong mean velocity.  Nonintrusive optical methods, such as laser-induced fluorescence (LIF), particle image velocimetry (PIV), and particle tracking are potentially viable alternatives. Particle tracking is best suited for our purposes, as it directly provides Lagrangian information without requiring interpolation or integration of velocity fields. But since particle tracking follows the motion of individual tracer particles, it has typically been restricted to fields that are sampled too sparsly to resolve the velocity gradient. At small Reynolds numbers, however, \citet{2005JFMLuthi} successfully used particle tracking to measure the velocity gradient; thus, if the seeding density can be made large enough, the gradient can be resolved. One promising method for increasing the seeding density in particle tracking is to illuminate not the entire measurement volume at once but rather to section it into successively illuminated slabs by scanning a laser through it~\citep{2005EIFHoyer}. This scanning particle tracking opens the door for using particle tracking to measure the velocity gradient at larger Reynolds numbers, and has been used, for example, along with LIF to study the joint evolution of the velocity gradient and the density field in turbulent gravity currents \citep{2014MSTKrug}.

In this paper, we report experimental results on the joint dynamics of rods and the velocity gradient tensor in turbulence using a scanning particle tracking system. In \S \ref{sec:example}, we show an example of a rod trajectory along with the local velocity gradient tensor to give a qualitative idea of the different ways the gradient contributes to the rod tumbling rate. In \S \ref{sec:setup}, we present in detail our experimental setup and our scanning particle tracking system; an analysis of our experimental uncertainties in measuring the gradient are reserved for the appendices. Details of the data analysis are addressed in \S \ref{sec:analysis}. In \S \ref{sec:results}, we discuss our experimental results concerning the dependence of the rod tumbling rate on the velocity gradient; in particular, we describe quantitatively the relative contributions from the vorticity and the strain rate and show that both are necessary for understanding the orientational dynamics of rods.

\section{Example}
\label{sec:example}

\begin{figure}
\begin{center}
$\begin{array}{cc}
\includegraphics[width=5.4in]{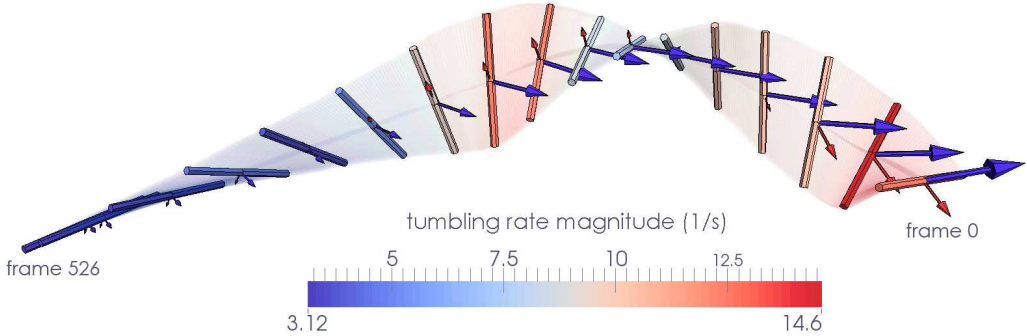}
\end{array}$
\caption{Reconstructed 3D trajectory of a rod (700 $\mu$m in length) over 1 sec (527 frames). The colormap shows the tumbling rate of a rod. The blue and the red arrows represent the vorticity vector ${\bf{\omega}}$ and the largest stretching direction $\hat{\bf{e}}_1$. The length of two arrows indicate the magnitude of ${\bf{\omega}}$ and the eigenvalue $\lambda_1$ of strain rate tensor corresponding to the largest stretching direction $\hat{\bf{e}}_1$.}
\label{fig:3d_track}
\end{center}
\end{figure}

Before describing our techniques for measuring rod motion and the velocity gradient tensor in detail, we show in figure \ref{fig:3d_track} an example of a measured rod trajectory along with vectors characterizing the local velocity gradients. The ribbon shows the full trajectory with a solid rod plotted only once every 25 time steps. The color indicates the magnitude of rod tumbling rate $|\dot{\bf{p}}|$, which depends on the straining and swirling motion of the surrounding flow. The tumbling rate due to strain is 
\begin{equation}
({\dot{p_S}})_i=\frac{\alpha^2-1}{\alpha^2+1}(S_{ij}p_j-p_ip_kS_{kl}p_l)
\end{equation}
and this tends to align the rod with the most extensional direction of the local flow (red arrow), which is given by the eigenvector $\hat{\bf{e}}_1$ of the strain-rate tensor $S_{ij}$ that corresponds to its largest eigenvalue $\lambda_1$. Local swirling is characterized by the rotation-rate tensor $\Omega_{ij}$, which tends to rotate the rods about the local vorticity direction $\hat{\bf{\omega}}$ (blue arrow) at a rate of 
\begin{equation}
(\dot{{{p}}_{\Omega}})_i=\Omega_{ij}p_j.
\end{equation}

\begin{figure}
\begin{center}
\includegraphics[width=5.4in]{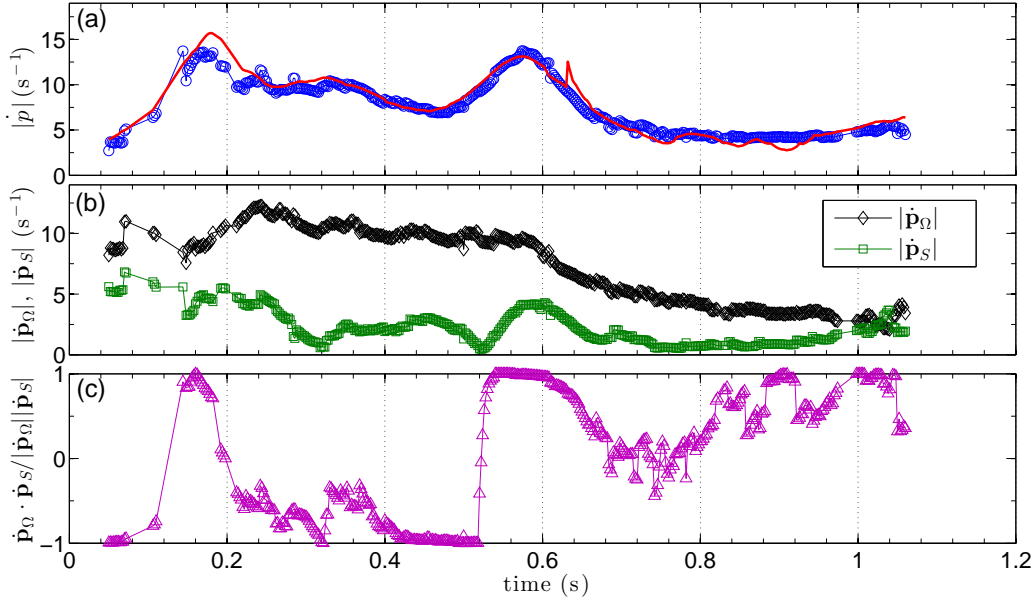}
\caption{(Color online) The time series of the rod tumbling rate for the same trajectory shown in figure \ref{fig:3d_track}. (a) The magnitude of total tumbling rate from two different measurements: $\dot{\bf{p}}$ determined by differentiating the rod orientation (red curve) and $\dot{\bf{p}}^J$ calculated from Jeffery's equation  using the velocity gradient measurements around the rod (blue circles). (b) Two contributions of tumbling rate from vorticity $|\dot{\bf{p}}_{\Omega}|$ and from strain $|\dot{\bf{p}}_S|$. (c) The cosine angle between two vectors $\dot{\bf{p}}_{\Omega}$ and $\dot{\bf{p}}_{S}$.}
\label{fig:rot_rate}
\end{center}
\end{figure}

For this rod trajectory, the magnitude of the total tumbling rate $\dot{{\bf{p}}}$ as well as its two component contributions $\dot{{\bf{p}}_{\Omega}}$ and $\dot{{\bf{p}}_S}$ are shown in figure \ref{fig:rot_rate}(a) and (b). The total tumbling rate can be computed in two ways: by differentiating the rod orientation signal or by using measurements of the velocity gradient tensor and Jeffery's equation $\dot{\bf{p}}^{J}=\dot{\bf{p}}_{\Omega}+\dot{\bf{p}}_S$. As shown in figure \ref{fig:rot_rate}(a), the two measurements agree well with each other, indicating both that our measurement of the velocity gradient tensor is accurate and that the rod is small enough that Jeffery's equation holds. In figure \ref{fig:rot_rate}(c), we plot the cosine of the angle between $\dot{{\bf{p}}_{\Omega}}$ and $\dot{{\bf{p}}_S}$. When this quantity is negative, the contribution to the rod tumbling due to strain works against that due to rotation. But when it is positive, the two contributions work cooperatively and lead to large tumbling rates, as can be seen at 0.16~s and 0.56~s in figure \ref{fig:rot_rate} (corresponding to the red regions in figure \ref{fig:3d_track}.

\section{Experimental apparatus and setup}
\label{sec:setup}

We generated a turbulent flow in an octagonal Plexiglass water tank measuring $1\times1\times1.5$ m$^{3}$. Two grids with a mesh size of 8 cm were oscillated in phase with an amplitude of 12~cm.  Details about the apparatus are given in ~\citet{blum2010}.  In this paper, all experiments were performed with a grid frequency of 1 Hz which creates turbulence in the center of the tank with $R_{\lambda}=140$. The Kolmogorov length scale is $\eta = (\nu^3/\epsilon)^{1/4}= 0.31$~mm and the Kolmogorov time scale is $\tau_{\eta}= (\nu/\epsilon)^{1/2}=0.093$~s, where $\nu$ is the kinematic viscosity and $\epsilon$ is the energy dissipation rate per unit mass whose measurement is described in Appendix B. The temperature of the water was almost constant at $21.8\pm 0.2^{\circ}$C, giving a kinematic viscosity of $\nu=(9.61\pm0.05)\times10^{-7}$~m$^2$/s. 

\begin{figure}
\begin{center}
$\begin{array}{cc}
\includegraphics[width=5.4in]{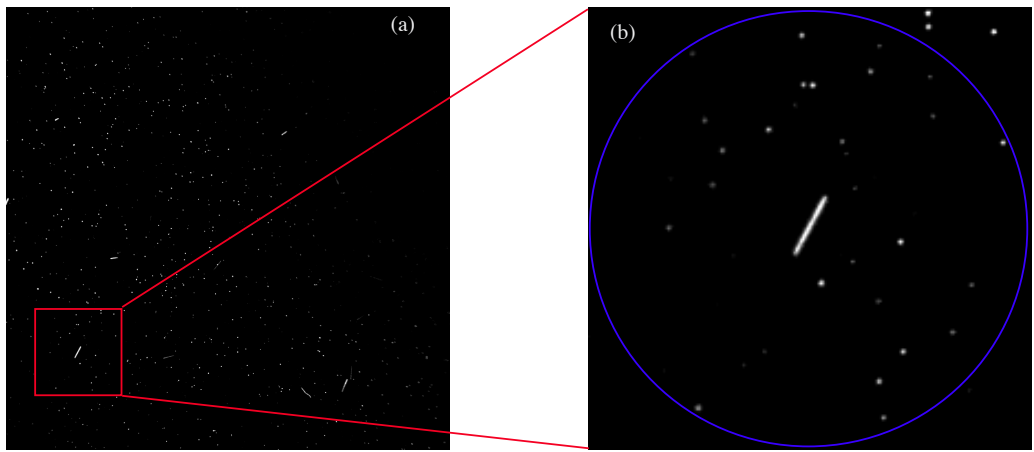}
\end{array}$
\caption{The full resolution (a) and zoomed-in (b) images captured by one of the three cameras. The blue circle is centered at the centroid of a rod with 2 mm (6.7 $\eta$) radius. The tracer particles that fall in such a sphere will be used to calculate the velocity gradient tensor.}
\label{fig:image}
\end{center}
\end{figure}

To measure the velocity gradient tensor simultaneously with rod motion, we need two kinds of particles: the rods themselves and small, spherical tracer particles. A key to this experiment is obtaining suitable particles. Fluorescent particles are convenient, as they will produce much better images, since some small residue from aluminum top and bottom plates and the bearings may be carried into the measurement volume.  For tracer particles, we therefore used internally dyed polystyrene divinylbenzene (PS-DVB) particles with 30~$\mu$m diameter and density 1.05 g/cm$^3$, purchased from Thermo Scientific. The rods were nylon fiber from DonJer Corp., with major and minor axes of roughly 700 and 30~$\mu$m, respectively, giving an aspect ratio of $23.3$. We dyed the rods with Rhodamine-B, so that they absorbed green light (wavelength $\lambda=532$~nm) from our laser and emitted red light at the same wavelength as the tracer particles. A typical image of these fluorescent particles, captured through a Schneider B+W MRC Orange 550 band-pass filter, is shown in figure \ref{fig:image}. The shapes of the two types of particles are well-defined and distinct from each other, and thus the particles can be easily separated using automated image analysis. Note that the brightnesses of the two kinds of particles are very similar, another important factor in the experiment: if the brightnesses of two types of particles were very different, it would be difficult to determine their positions accurately at the same time. Even though the tracer particles are much smaller than the rods, the internal dying emitted a much stronger fluorescent signal than the rods, which we dyed ourselves. 

In our subsequent analysis, we make the assumption that the rods and the tracers do not exhibit inertial effects caused by their finite size or density difference with respect to the fluid. To characterize the validity of this assumption, we use the Stokes number $\mathrm{St}=\tau_p/\tau_{\eta}$, given by the ratio between the time scale of the Stokes viscous drag $\tau_p=r^2/(3\beta\nu)$ and the Kolmogorov time scale $\tau_{\eta}$. Here, $r$ is the radius of the particle and $\beta=3\rho_f/(2\rho_p+\rho_f)$ is a coefficient capturing the effect of a density difference between the fluid (with density $\rho_f$) and the particle (with density $\rho_p$). If the particle response time $\tau_p$ is much smaller than the smallest scale $\tau_{\eta}$, the inertial effect of the particle is negligibly small, and the particles can be safely treated as tracers. For our spherical particles, $\mathrm{St}=8.6\times10^{-5} \ll 1$. For a rod, the relaxation time is given by $\tau_r=2\alpha \rho_pr^2ln(\alpha+\sqrt{\alpha^2-1})/9\nu\sqrt{\alpha^2-1}\rho_f$, where $r$ is the semiminor axis \citep{Zhang2001,Shapiro1993}. Using this expression, the Stokes number for the rods is $\tau_r/\tau_p=8\times10^{-3}$, which is also much less than unity. Thus, both the rods and the spherical particles are in the tracer limit, and will follow the fluid motion accurately.

\begin{figure}
\begin{center}
\includegraphics[width=4.0in]{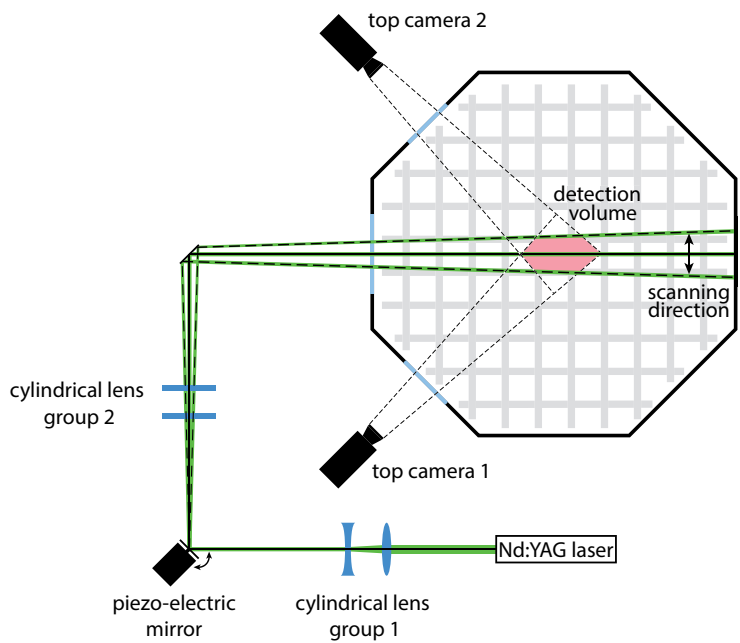}
\caption{Schematic of the experimental apparatus. Lens group 1 controlled the thickness of the laser beam and needed to be placed before the piezo-electric mirror. Lens group 2 expanded the beam in the vertical direction to form the final laser sheet. The laser sheet stayed nearly parallel across the whole scanning range (divergence is exaggerated in the figure). Detection volume is not to scale. Camera 3 is located below camera 1.}
\label{fig:scanning}
\end{center}
\end{figure}

\subsection{Scanning System}

Once the particles are chosen, we need to determine the proper seeding density for the experiment. Generally speaking, we want a very high density of spherical particles and relatively low density of rods. The density of spherical particles is directly related to the spatial resolution, which is crucial for measuring the velocity gradient tensor. The maximum density is limited, however, if we illuminate the entire measurement volume, because the images of individual particles will overlap with each other when the particle seeding density is high. We therefore use a scanning particle-tracking system for our measurements \citep{2005EIFHoyer}. The basic principle of this technique is a sacrifice of temporal resolution for improved spatial resolution. By subdividing the measurement volume into 10 slabs, for example, and successively scanning the illumination through them, we can increase the particle seeding density by a factor of 10 (in the ideal case), albeit at the cost of requiring a factor of 10 increase in camera frame rate and a decrease in total recording time.

A schematic of the experimental setup is shown in figure \ref{fig:scanning}. The beam from a pulsed Nd:YAG laser with an average output power of 50~W was stretched independently in height and in width by two sets of lens to create an illumination slab measuring 50~mm by 3~mm. To ensure a relatively uniform slab depth over a height of 30~mm, we did not image the top and bottom 10~mm of the laser slab. The scanning of the light slab was controlled by a piezo-electric driven mirror with a diameter of 15~mm and a maximum deflection angle of 2.2~mrad with sub-$\mu$rad resolution. The small deflection angle is magnified through the optical system to give a 12~mm scanning range in the center of the flow chamber. For reasons we will explain in Sec \ref{sec:pt}, each slab overlaps with the previous one by $\sim 50$\%. Compared to previous designs using a rotating prism to scan the illumination slab \citep{2005EIFHoyer}, a piezo-electric driven mirror or an acousto-optic deflector has the potential to generate faster scanning rates, which are more suitable for turbulence with even higher Reynolds number.

Three Photron FASTCAM SA5 cameras with a resolution of $1024 \times 1024$ pixels were used to image the particles in a small volume measuring approximately $3 \times 3 \times 4$~cm$^{3}$ in the center of the tank. To resolve this small volume more than a half meter away from the side walls of the tank, each camera was fitted with a Nikkor 200 mm macro lens and a Kenko 1.6 teleconverter. The cameras were mounted to a custom-built frame on an optical table uncoupled from the turbulence tank to minimize camera vibration. Two of the cameras (labelled as ``top cameras'' in figure \ref{fig:scanning}) were mounted in the same lateral plane with a 90$^{\circ}$ angular separation, and looked down into the detection volume at an angle of 16$^{\circ}$. The third camera was aimed up toward the same volume at an angle of 21$^{\circ}$. Situating the three cameras in different planes helps to increase the stereomatching accuracy \citep{2006EIFOuellette}. 

The piezo-electric mirror was driven with an adjusted saw-tooth signal at a frequency of 500~Hz, rising linearly for 80\% of each cycle and falling for the remaining 20\%. This driving produced a nearly linear scanning motion of the laser beam through the measurement volume followed by a quick return to the initial position. The cameras recorded images at a frame rate of 5000~Hz, so that each cycle of the mirror resulted in 10 captured sets of images, with 8 of these linearly positioned through the measurement volume. We set the cameras to store only 8 out of every 10 frames to conserve on-board memory, which could hold a total of 5456 images. Transferring the contents of the camera memory to a computer hard drive took several minutes. Our experimental protocol thus consisted of several steps. First, the two grids were driven at 3~Hz for 30 seconds to stir up the fluid and particles. The grids were then slowed to a 1~Hz oscillation rate, and the flow was given 1 minute (approximately 17.5 large-eddy turnover times) to stabilize. The cameras then recorded images until their memory was full (for about 1 second), and the system subsequently rested until the data transfer to the hard drive was complete. The timing of this system was automatically controlled with Labview scripts. The results reported here come from a full day of measurements, giving roughly 300 data sets. 

\begin{figure}
\begin{center}
\includegraphics[width=2.5in]{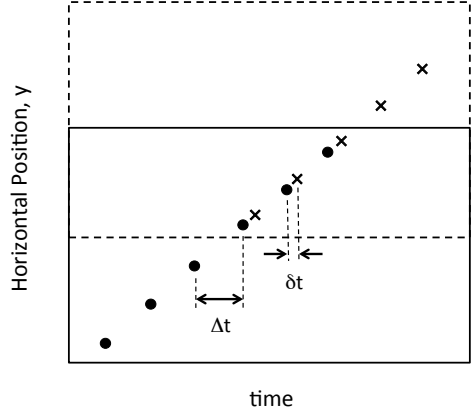}
\caption{Diagram of one particle passes from one slab $n$ (box in solid line) to the next one $n+1$ (box in dashed line). The positions of the same particle in slab $n$ and $n+1$ are represented by solid circles and crosses, respectively. The time between two consecutive volume scans is $\Delta t$, and the time between two slabs is $\delta t$. In the overlapping region, there are three pairs of joins, which can be used to connect two segments of trajectories in different slabs together.}
\label{fig:track}
\end{center}
\end{figure}

\begin{figure}
\begin{center}
$\begin{array}{cc}
\includegraphics[width=5.4in]{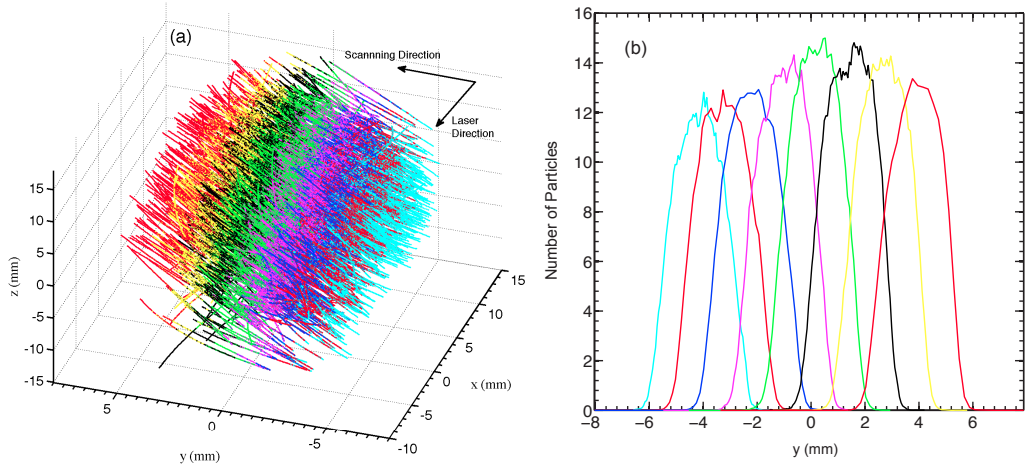}
\end{array}$
\caption{(Color online) (a) The trajectories that pass through at least five laser slabs in one typical movie with 5457 frames. Particle positions that belong to different laser slabs are shown by different color. (b) The histogram of y positions of the particles in each slab.  Only particles from the center of the detection volume (-5 mm$<x<$5 mm and -5 mm$<z<$5 mm) were used to ensure that the histogram represents the slab width and overlap. }\label{fig:sample}
\end{center}
\end{figure}

\subsection{Simulation}

To compare with the experimental data, one dataset from a direct numerical simulation (DNS) of homogeneous isotropic turbulence at Reynolds number $R_{\lambda}=180$ was used. There are total $N^3=512^3$ collocation points for the entire volume. A total of $7\times 10^4$ Lagrangian trajectories of the velocity gradient tensor were followed for $O(1)$ large-eddy turnover times, and the time step for integrating the Navier--Stokes equations and tracking Lagrangian points was $O(10^{-2}\tau_{\eta})$. Along each Lagrangian trajectory, the orientation of a virtual infinitesimal rod with an aspect ratio of $20$ was computed by integrating Jeffery's equation (equation \ref{eq:Jeff}) using a fourth-order Runge--Kutta method \citep{2012PRLShima}. The details of this simulation are given by \citet{2009PREBenzi}. 

\section{Data Analysis}
\label{sec:analysis}
\subsection{Image processing}
To determine particle positions, the digital image from each camera is first segmented into groups of bright pixels, representing both, tracers particles and rods. Typically, the image of one spherical tracer particle contains $4\sim 9$ pixels, corresponding to 2 to 3 pixels in diameter. The number of pixels in a rod image depends on its relative orientation to the camera. The maximum size is $\sim 320$ pixels when its long axis is perpendicular to the optical path of the camera, in which case it is also highly anisotropic. But the image of a rod becomes quasi-spherical with an area of $\sim 16$ when it points directly toward the camera. Ideally, the images of rods and spherical particles could be separated solely by their area. However, sometimes a spherical particle looks larger if it is out of focus. So we add another criterion for separating the images of spheres and rods: eccentricity, which is determined by finding the best fit ellipse to the pixel cluster and calculating the ratio of the distance between the foci and the major axis of the ellipse. We find that the eccentricity ranges from 0 to 0.6 for spherical particles and from 0.8 to 1 for rods. In practice, a particle image is considered to be a rod if its eccentricity is larger than 0.9 and its area is larger than 30 pixels. These criteria separate almost all rods from spherical tracer particles. Sometimes a rod that points directly toward one camera will be mistakenly identified as a tracer in that camera; but because we have other cameras viewing the same rod from different directions, where it will be distinctly elongated, it can be correctly identified by combining the information from all the cameras.

\subsection{Particle Tracking}
\label{sec:pt}
Once we know the particle type, its center is determined by averaging the positions of its pixels weighted by their brightness values. This procedure is used for both types of particles. From the two-dimensional (2D) positions of the particles as measured by each camera, the three-dimensional (3D) positions can be found by stereoscopic matching, and subsequently connected together in time from one volume scan to the next \citep{2006EIFOuellette}, although some modifications of common tracking algorithms are required. 

In traditional particle tracking, the entire measurement volume is illuminated and is imaged at a constant rate of $1 / \Delta t$, where $\Delta t$ is the time interval between two frames. Thus, finding candidate particles to extend a given trajectory requires searching for potential matches only at a time $\Delta t$ in the future. But for the scanning system, as is sketched in figure \ref{fig:track}, there are \emph{two} relevant time intervals: $\Delta t=1/500$~s, the time between two full scans of the volume, and $\delta t=1/5000$~s, the time between the illumination of two neighboring slabs of the volume. Since a particle may or may not pass from one slab to another over an interval of $\delta t$, the time at which a particle corresponding to the continuation of a trajectory may be found is not obvious: it may be found $\delta t$ in the future, for example, if the particle moved between two successive slabs, at $\Delta t$ in the future if it remained in the same slab, at $\Delta t - \delta t$ in the future if it moved to a \emph{previous} slab, or even potentially at other times. A previous scanning tracking system by  \citet{2005EIFHoyer}, performed the search for the continuation of a particle's trajectory in slabs n, n-1, and n+1. To handle this ambiguity, we first track the particles found in each individual slab separately, and then subsequently merge the short segments that belong to different slabs but that refer to the same particle. Within each slab, particles are tracked using a standard predictive tracking algorithm \citep{2006EIFOuellette}, with a recording time of roughly $\sim d/\Delta t\tilde{u}=$ 75 frames, where $d=3$~mm and $\tilde{u}\sim2$~cm/s are the thickness of one laser slab and the root-mean-square velocity of the flow, respectively. In figure \ref{fig:sample}(b), it is seen that each pair of neighboring slabs overlaps by $d_{ov}\sim 1.5$~mm. Thus, since the velocity of a particle is bounded, a particle that moves from one slab to the next will inevitably pass through the overlapping region and will be recorded twice during each volume scan. We refer to each doubly recorded position as a ``join'' between the two trajectory segments. In figure \ref{fig:track}, where we schematically demonstrate our tracking method, there are four joins. In principle, two joins are enough to merge two trajectory segments into one longer track. Estimating the number of joins in the overlapping region using simple kinematics gives $d_{ov}/\Delta t\tilde{u}=32$. Thus, even a particle with a speed of $10\tilde{u}$ would still have more than two joins in the overlapping region. We apply this merging procedure consecutively for all neighboring slabs, and thus link all segments in different slabs together into longer trajectories. In figure \ref{fig:sample}(a), we show those trajectories that pass through at least 5 different slabs, which evenly cover the entire volume, to demonstrate qualitatively that our measurements are robust and that the overlapping regions are large enough for splicing. 


The stereomatching and tracking procedures for rods are almost the same as for the tracers. The primary difference is that we need to keep track of the orientation of a rod in addition to its position. In 2D, the orientation, defined as the angle between the major axis of the rod and horizontal axis of the image, can be extracted from all three cameras. The 3D orientation of rods can then be uniquely determined from these three 2D angles and the viewing directions of the cameras \citep{2013thesisShima}. Errors in determining the orientation arise mainly from the finite aspect ratio of the rods. In the experiments reported here, the aspect ratio of the rods is 20, roughly 4 times that in our previous experiments \citep{2012PRLShima}. Thus, the uncertainty in orientation in these data is smaller than it was for our previous results \citep{2013thesisShima}.

\subsection{Interpolation and differentiation}
The next step in processing the data is to filter and differentiate the trajectories to obtain the velocities of the tracers and tumbling rates of the rods. Common methods to accomplish this task include convolving the trajectory with appropriate kernels \citep{2004PDMordant} and polynomial fitting \citep{2002JFMVoth}. In the scanning system, the convolution method is difficult to apply because the points along the trajectories are not evenly spaced in time, leading to difficulties in accurate numerical integration. We therefore use polynomial fitting. In addition to smoothing and differentiating, fitting also allows us to accurately interpolate the measured positions and velocities in time so that all data throughout the measurement volume is contemporaneous. That is, we can use the fits to extract the kinematics of the flow field not at the measured space-time positions of the particles but rather at the times $t_n=4\delta t+n\Delta t$ ($n=0,1,2,3,\ldots$), so that we acquire one full velocity field for each scan of the volume, measured at the time corresponding to the illumination of slab 5 (halfway through the volume scan). To accomplish this interpolation, for each $n$, we fit a polynomial to all measured data points in the range $[t_n-(\tau_f-1)/2,t_n+(\tau_f-1)/2]$ along each track, where $\tau_f=9\Delta t$ is the temporal length of the fit. This choice minimizes noise without unduly affecting the signal \citep{2002JFMVoth}. From the polynomial fits, we extract smoothed positions, velocities, and accelerations of the tracers at $t_n$ spread over the entire volume. We apply a similar process to the rods to extract their orientation and tumbling rate. Note, however, that we define the orientation of the rods by a unit vector along their major axis. Smoothing the orientation signal decreases random error in the measurement of the orientation direction, but may change the magnitude of this vector. Therefore, we must re-normalize the orientation vector for each rod after smoothing.

\subsection{Seeding density and shape factor}
To determine how large we could make the seeding density of the spherical tracer particles and still obtain good measurements, we tested the scanning system by slowly increasing the number of tracer particles. When the particle concentration is low, the ratio between the number of successfully stereomatched particles to the number of particles detected in each 2D image is almost constant. We can determine the maximum seeding density at which we can still resolve the particles by locating the point at which this ratio begins to decrease. In our experiments, this point corresponds to roughly 500 stereomatched particles in each slab. After accounting for doubly imaged particles in the overlapping regions and trajectories shorter than $\tau_f$, for which we cannot measure velocities or accelerations, we can reliably measure about 2000 velocity vectors in each volume scan. This number varies somewhat over time due to sedimentation of particles to the sidewalls and bottom plate, so we add particles over the course of an experimental run to maintain a roughly constant particle concentration. The seeding density of rods is kept low to avoid interactions between them. We have roughly 10 rods in the measurement volume at any given time, so the non-dimensional concentration is roughly $n \ell^3 \sim 10^{-3}$, where $n$ is the number density and $\ell$ is the length of a rod.  This is far below the concentration at which rod-rod interactions become important.  

After tracking, smoothing, and differentiation, we are left with trajectories of rod orientations along with the velocities of many tracers surrounding them. Measuring the velocity gradient tensor around each rod requires us to estimate the spatial gradient from multiple velocity vectors inside a small volume. To do this, we first locate all tracers within a 2~mm ($\sim 6\eta$) radius of the center of a rod. This radius is chosen to have sufficient tracer particles ($6\sim10$, given our tracer-particle seeding density) surrounding the rod to estimate the velocity gradient well. This number is comparable to what has been used in previous experiments, where a radius of nearly $8\eta$ was found to be sufficient so that viscous effects dominate and the velocity field is close to linear \citep{2005JFMLuthi}. 

To estimate the gradient, consider $N$ tracers at positions ${\bf{x}}^n(t)$ and with velocities ${\bf{u}}^n(t)$ ($n=1,2,3,\ldots,N$) that are sufficiently close to a rod. Their relative position and velocity with respect to the center of mass of the particle cloud are ${\bf{x}}'^{n}(t)={\bf{x}}^n(t)-\sum_n{\bf{x}}^n(t)/N$ and ${\bf{u}}'^{n}(t)={\bf{u}}^n(t)-\sum_n{\bf{u}}^n(t)/N$, respectively. Determining the velocity gradient tensor $A_{ij}$ can then be formulated as a least-squares problem by finding the minimum value of the squared residuals $S=\sum_n[u_i'^{n}(t)-A_{ij}x'^{n}_j(t)]^2$ \citep{2013POFAlain}. In general, however, the tracer particles in the cloud surrounding a rod are randomly distributed in space, and will sometimes lie in almost in the same plane. Such cases will introduce a large error in the determination of the out-of-plane components of the velocity gradient. To exclude these cases, we use the inertia tensor ${\bf{I}}={\bf{g}}/tr(\bf{g})$ with $g_{ij}=\sum_n x_i^nx_j^n$ to characterize the shape of the tracer-particle cloud. $\bf{I}$ can be diagonalized in an orthogonal basis with three eigenvalues $I_1\leq I_2\leq I_3$. For a symmetric object, $I_1= I_2= I_3=1/3$, while for a co-planar particle cloud $I_3\approx0$. In our experiment, we require $I_3>0.1$ to rule out cases that will have large errors. Empirically, we find that our results are not sensitive to the choice of this threshold when it is in the range from 0.07 to 0.2. In addition to the overall shape of the particle cloud, the distribution of the tracer particles inside the cloud may also affect the estimate of the velocity gradient: sometimes, the particles will all be concentrated in a corner, for example, and the center of mass of the tracer cloud will be far away from the rod center. This situation will lead to a biased estimate of the velocity gradient at the rod center. To avoid this case, we require that the distance between the rod center and center of the tracer cloud be no more than one third of the radius of gyration of the tracer cloud. 

\subsection{Alignment between vorticity and strain rate tensor}
\begin{figure}
\begin{center}
$\begin{array}{cc}
\includegraphics[width=5.4in]{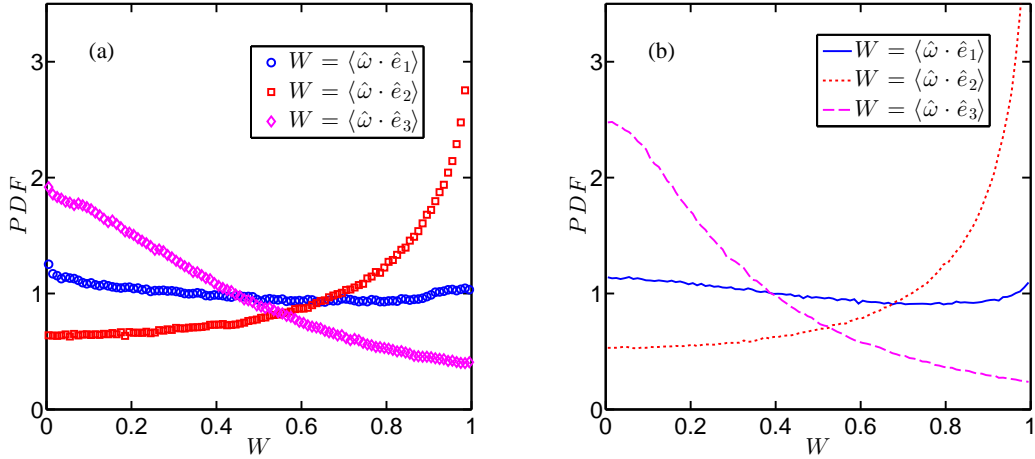}
\end{array}$
\caption{(Color online) Probability distribution function of the cosine of the angle between vorticity $\hat{\bf{\omega}}$ and with eigenvectors of the strain-rate tensor $\hat{\bf{e}}_i$ (a) experimental measurements (b) simulation results}
\label{fig:wei}
\end{center}
\end{figure}

Details of the errors in our measurements and a comparison between the dissipation rate estimated from the measured velocity gradient tensor and the velocity structure functions are reported in Appendices A and B. To demonstrate the quality of our measurements, however, we briefly consider some of the well-known geometric properties of the velocity gradient - in particular, the alignment between the vorticity and the eigenvectors of the rate of strain. The rate of strain $S_{ij}$ is a symmetric second-rank tensor, and can be described by its three eigenvectors $\hat{\bf{e}}_i$ ($i=1,2,3$), which correspond to the eigenvalues $\lambda_i$ with $\lambda_1\geq\lambda_2\geq\lambda_3$. Intuitively, one would expect that the vorticity $\hat{\bf{\omega}}$ would tend to align with the most extensional eigenvector, $\hat{\bf{e}}_1$. But, in general, vorticity is preferentially aligned with the intermediate eigenvector $\hat{\bf{e}}_2$ \citep{1987POFAshurst}.  We show the PDFs of the cosine of the angle between $\hat{\bf{\omega}}$ and the $\hat{\bf{e}}_i$ in figure \ref{fig:wei} for our measured velocity gradients and for the direct numerical simulation. The overall trend of the PDFs is very similar for the experiments and the simulation: as expected, the vorticity is best aligned with $\hat{\bf{e}}_2$. The trends, however, are more pronounced in the simulation due to experimental inaccuracies in the measurement of the velocity gradient, and potentially from the coarse-graining introduced by estimating the velocity gradient from a finite-sized tracer cloud. 

\section{Results and Discussion}
\label{sec:results}

\begin{figure}
\begin{center}
$\begin{array}{cc}
\includegraphics[width=5.4in]{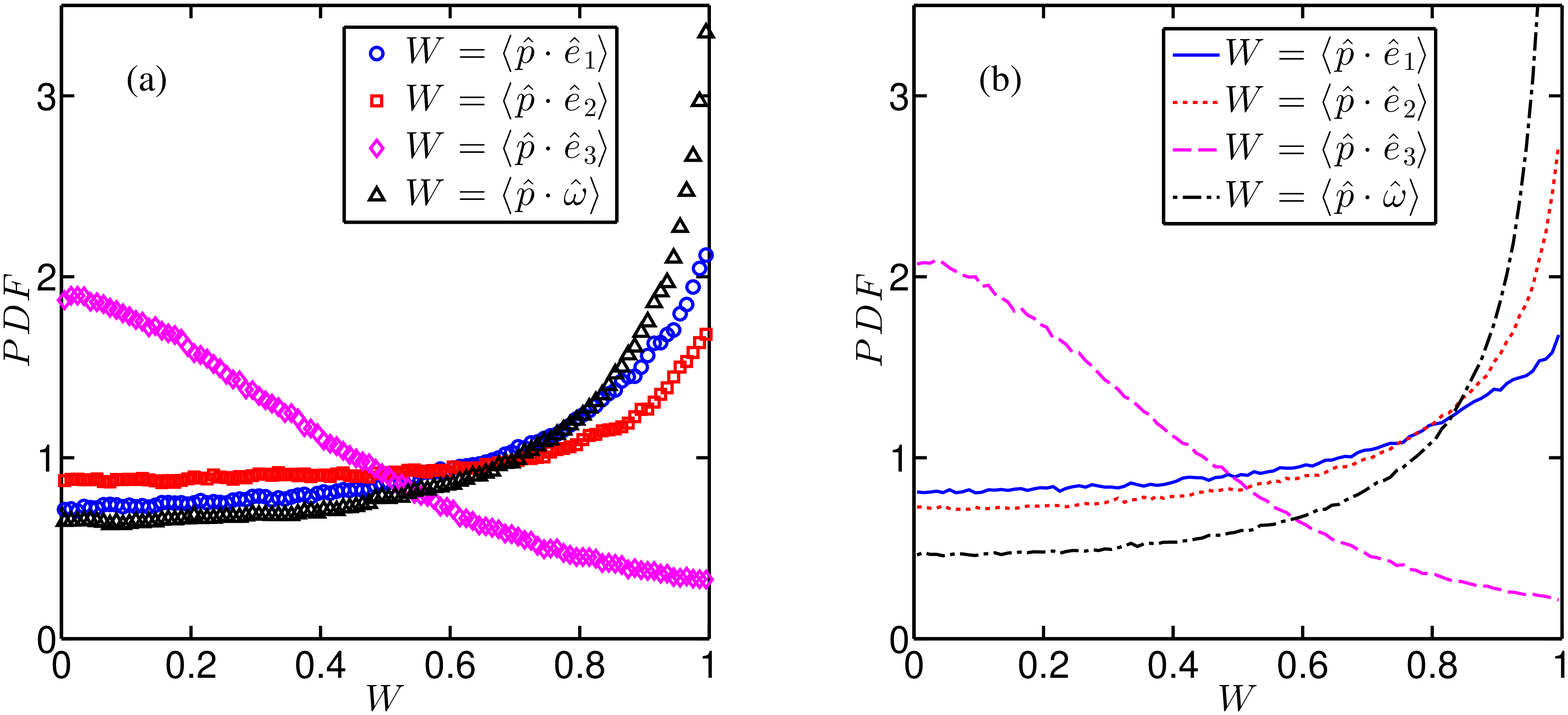}
\end{array}$
\caption{(Color online) Probability distribution function of the cosine of the angle between the rod orientation $\hat{\bf{p}}$ with vorticity $\hat{\bf{\omega}}$ and eigenvectors of the strain-rate tensor $\hat{\bf{e}}_i$ (a) experimental measurements (b) simulation results}
\label{fig:pei}
\end{center}
\end{figure}

\begin{figure}
\begin{center}
$\begin{array}{cc}
\includegraphics[width=2.4in]{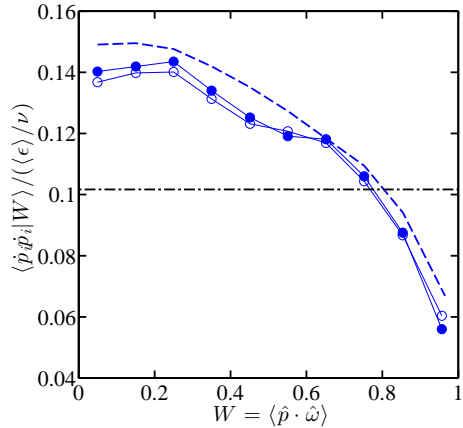}
\end{array}$
\caption{(Color online) Mean squared tumbling rate of rods conditioned on the alignment between rod and vorticity $\langle\hat{\bf{p}}\cdot\hat{\bf{\omega}}\rangle$. The black dash-dotted line represents the mean squared tumbling rate and dashed line is from simulation results. Open symbols show the tumbling rate from differentiating rods' trajectories of orientation, and closed symbols show the calculation from Jeffery's equation applied to the measured velocity gradient tensor.}
\label{fig:condpw}
\end{center}
\end{figure}

\begin{figure}
\begin{center}
$\begin{array}{cc}
\includegraphics[width=5.4in]{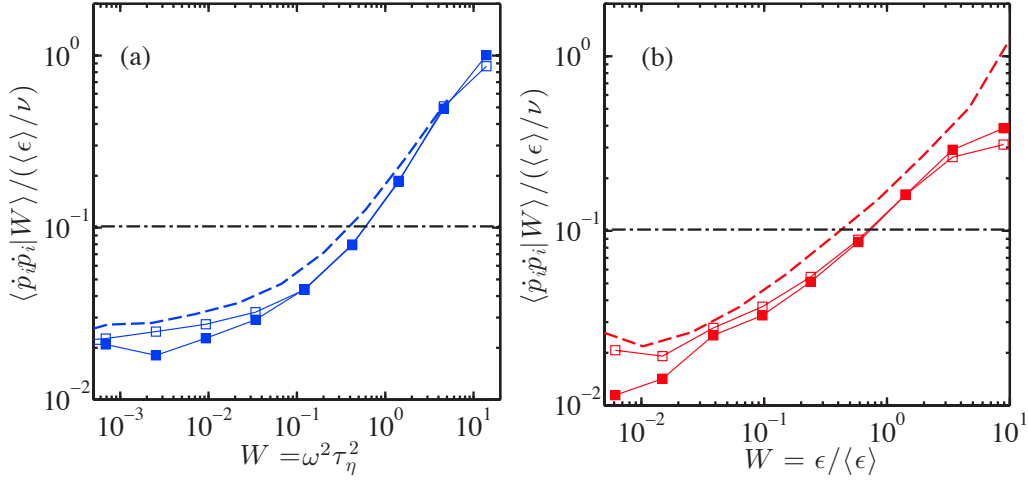}
\end{array}$
\caption{(Color online) Mean squared tumbling rate of rods conditioned on (a) normalized enstrophy (b) normalized dissipation rate. The symbols are the same as they are in figure \ref{fig:condpw}.}
\label{fig:rotrweps}
\end{center}
\end{figure}

\begin{figure}
\begin{center}
$\begin{array}{cc}
\includegraphics[width=5.4in]{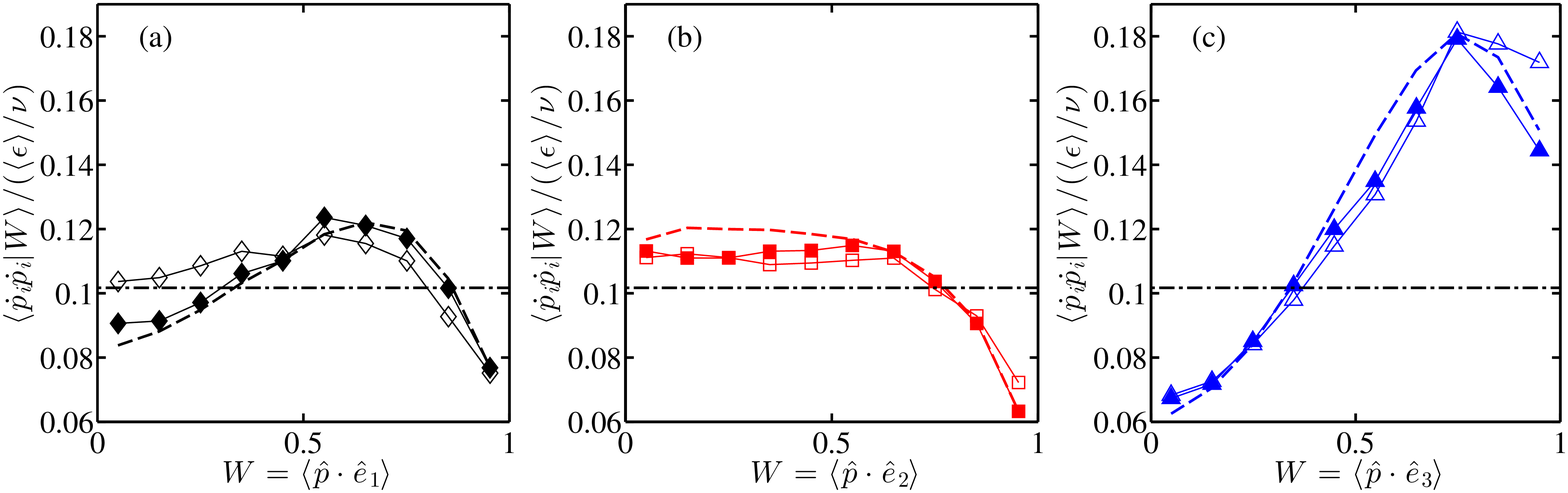}
\end{array}$
\caption{(Color online) Mean squared tumbling rate of rods conditioned on the alignment between rod and eigenvectors of strain rate tensor (a) $\langle\hat{\bf{p}}\cdot\hat{\bf{e}}_1\rangle$ (b) $\langle\hat{\bf{p}}\cdot\hat{\bf{e}}_2\rangle$ (c) $\langle\hat{\bf{p}}\cdot\hat{\bf{e}}_3\rangle$. The symbols are the same as they are in figure \ref{fig:condpw}.}
\label{fig:condpei}
\end{center}
\end{figure}

Figure \ref{fig:pei} shows the alignment of rods with respect to the vorticity, $\hat{\bf{\omega}}$, and the three eigenvectors of the strain rate, $\hat{\bf{e}}_i$,  for both our experiments and the simulation. For both cases, rods are much more strongly aligned with $\hat{\bf{\omega}}$ than they are with $\hat{\bf{e}}_i$. This is because $\hat{\bf{\omega}}$ and $\hat{\bf{p}}$ each independently tend to align with the strongest Lagrangian stretching direction, as defined by the maximum eigenvector of the left Cauchy--Green strain tensor \citep{2014JFMNi}.   The alignments of rods with the strain rate eigenvectors are weaker because rods align with the Lagrangian stretching integrated over several Kolmogorov times, and this direction is usually quite different from the instantaneous stretching direction defined by the strain rate eigenvectors.  

The alignment distributions from the simulation in Figure \ref{fig:pei} are in excellent agreement with previous simulations~\citep{2011NJPPumir}.   Our experiment agrees quite well with the simulations, except that the experiments show that rods are slightly better aligned with $\hat{\bf{e}}_1$ than they are with $\hat{\bf{e}}_2$, while the simulations show the opposite. Prior work on material lines \citep{2008thesisWan} has shown that the relative strength of their alignment with $\hat{\bf{e}}_1$ and $\hat{\bf{e}}_2$ is sensitive to Reynolds number and perhaps to the particular driving of the flow. Material lines rotate just like thin rods with large aspect ratio $\alpha\gg1$. Thus, the small difference between the experiment and the simulation may result from differences in Reynolds number and the forcing mechanism.

In turbulence, the tumbling rate of rods has been found to be much smaller than it would be if the rods were randomly oriented \citep{2005Shin,2012PRLShima}.
This decrease has been qualitatively understood to be a result of the alignment of rods with the vorticity vector.    Given the alignment of rods with the vorticity,  rods will preferentially rotate about their long axis, and therefore the contribution of $\Omega_{ij}$ to the rod tumbling rate $\dot{\bf{p}}$ will be weak. 

To understand the rod tumbling rate in more detail, we can start with Jeffery's equation to estimate the parameter space and evaluate the complexity of the problem. Recall that the velocity gradient tensor ${\bf{A}}$ is a second-rank tensor with 8 independent components, assuming incompressibility.  For turbulent flow with isotropic small scales, the overall orientation in space does not matter so 5 scalars are sufficient to characterize ${\bf{A}}$ \citep{2011ARFMMeneveau}. There are many possible choices for these 5 scalars, such as two of the eigenvalues of ${\bf{S}}$ and the three components of the vorticity in the ${\bf{S}}$ eigenframe, or five scalars constructed from moments of the velocity gradient tensor such as the well known $R=-(A_{im}A_{mn}A_{ni})/3$ and $Q=-(A_{im}A_{mi})/2$ \citep{1992POFCantwell}. Specifying the relative orientation of the rod in the ${\bf{S}}$ eigenframe requires two additional independent angles. Thus, together with the five scalars required to determine ${\bf{A}}$, fully characterizing the rod tumbling rates requires a seven-dimensional parameter space. Denoting the generalized coordinates in this space by ${\bf{X}}({\bf{p}},{\bf{A}})$ and the corresponding probability density function as $\mathcal{P}({\bf{X}}({\bf{p}},{\bf{A}}))$, the mean-squared rod tumbling rate can be expressed as
\begin{equation}
\begin{split}
\langle\dot{p}_i\dot{p}_i\rangle=\int d{\bf{X}}({\bf{p}},{\bf{A}})\mathcal{P}({\bf{X}}({\bf{p}},{\bf{A}}))\left[\Omega_{ij}p_j+\frac{\alpha^2-1}{\alpha^2+1}(S_{ij}p_j-p_ip_kS_{kl}p_l)\right]
\\
\times\left[\Omega_{im}p_m+\frac{\alpha^2-1}{\alpha^2+1}(S_{im}p_m-p_ip_qS_{qn}p_n)\right]
\\
=\int \left[ \Omega_{ij}\Omega_{im}p_jp_m+\left(\frac{\alpha^2-1}{\alpha^2+1}\right)^2\left(S_{ij}S_{im}p_jp_m+ p_ip_kS_{kl}p_lp_ip_qS_{qn}p_n\right)+T_c\right]
\\
d{\bf{X}}({\bf{p}},{\bf{A}})\mathcal{P}({\bf{X}}({\bf{p}},{\bf{A}}))
\end{split}
\label{eq:seven}
\end{equation}
where $T_c$ represents six cross terms between strain and rotation. If we assume that the rods are randomly oriented and are uncorrelated with the velocity gradient tensor, this equation simplifies considerably because some of the averages can be taken independently; for example, $\langle S_{ij}S_{im}p_j p_m\rangle=\langle S_{ij}S_{im}\rangle\langle p_j p_m\rangle$. This assumption leads to 
\begin{equation}
\langle \dot{p}_i \dot{p}_i\rangle=\frac{\langle\Omega_{ij}\Omega_{ij}\rangle}{3}+\frac{1}{5}\left(\frac{\alpha^2-1}{\alpha^2+1}\right)^2\langle S_{ij}S_{ij}\rangle.
\end{equation}
Given that, in isotropic turbulence, $\langle S_{ij}S_{ij}\rangle=\langle\Omega_{ij}\Omega_{ij}\rangle=\langle\epsilon\rangle/\nu$, the normalized rod tumbling rate would depend only on its eccentricity:
\begin{equation}
\frac{\langle\dot{p}_i \dot{p}_i\rangle}{\langle\epsilon\rangle/\nu}=\frac{1}{6}+\frac{1}{10}(\frac{\alpha^2-1}{\alpha^2+1})^2.
\end{equation}

However, in turbulence, rods are \emph{not} randomly oriented, but are coupled with the velocity gradient tensor. Thus, characterizing their tumbling requires understanding of the full seven dimensional PDF $\mathcal{P}({\bf{X}}({\bf{p}},{\bf{A}}))$.   Since the Lagrangian velocity gradients are similar in many different turbulent flows, this PDF should be approximately universal.  Experimentally, we do not have enough samples to obtain this PDF at a reasonable bin size, and it would be difficult to present even if we obtained it.  The typical way to approach this complicated PDF would be to try to find a suitable low dimensional projection to simplify the problem. We will show below that although the most important two dimensions are the magnitude of the strain and the enstrophy, the relative orientation of a rod in turbulence cannot be neglected in determining its tumbling rate.

In the rest of this paper, we will use the conditional average of the rod tumbling rate along different dimensions to characterize the importance of each dimension and to provide new experimental insights into rod tumbling dynamics. Recall that, as shown in figure \ref{fig:rot_rate}, we have two different ways of experimentally determining the rod tumbling rate: a direct measurement, $\dot{p}_i$, and a measurement inferred from Jeffery's equation, $\dot{p}_i^J$. To check for any potential systematic offset between these different measurements, we computed the mean-squared tumbling rates for each, finding $\langle \dot{p}_i \dot{p}_i \rangle = 0.10$$\langle \epsilon \rangle/ \nu$ and $\langle \dot{p}_i^J \dot{p}_i^J \rangle = 0.09$$\langle \epsilon \rangle/ \nu$. The mean-squared tumbling rate computed from the simulation data was also $0.09$$\langle \epsilon \rangle / \nu$. These differences are very small; nevertheless, to compare the two, we systematically shifted the values of $\dot{p}_i^J$ and the tumbling rates computed from the simulation so that their mean-squared values were $0.10$. The shift was done by multiplying all data points in $\dot{p}_i^J$ and simulation results by 1.11, which will only move the curve up without changing its trend.

In figure \ref{fig:condpw}, we plot the mean-squared rod tumbling rate conditioned on $\langle\hat{\bf{p}}\cdot\hat{\bf{\omega}}\rangle$ for $\dot{p}_i$, $\dot{p}_i^J$, and the simulation. All three curves nearly collapse, and show that the rod tumbling rate monotonically decreases by more than 50\% as its alignment with the vorticity increases. For $\langle\hat{\bf{p}}\cdot\hat{\bf{\omega}}\rangle=0$, the rod is almost perpendicular to the vorticity vector and its tumbling rate has a larger contribution from the vorticity than from the strain.  If only the vorticity contributed, the mean-squared tumbling rate should be $\langle\dot{p}_i\dot{p}_i\rangle\approx\langle\Omega_{ij}\Omega_{ij}\rangle/3$. Since $\langle \Omega_{ij}\Omega_{ij}\rangle=\langle\epsilon\rangle/2\nu$, $\langle\dot{p}_i\dot{p}_i\rangle/(\langle\epsilon\rangle/\nu)$ at $\langle\hat{\bf{p}}\cdot\hat{\bf{\omega}}\rangle=0$ should be $1/6$, which is close to but slightly larger than the measured result of $\sim 0.14$. This discrepancy indicates that the strain contribution for $\langle\hat{\bf{p}}\cdot\hat{\bf{\omega}}\rangle=0$ is on average in the direction opposed to the vortical motion, just as it would be for a rod in Jeffery orbits in a uniform shear flow.

In the other limit, for $\langle\hat{\bf{p}}\cdot\hat{\bf{\omega}}\rangle=1$, the rod is almost perfectly aligned with the vorticity, so its tumbling rate has no contribution from vortical motion. The tumbling rate does not vanish, however, due to the coupling to the strain field. The contribution from the strain ${\bf{S}}$ can be investigated by considering the alignment of its three orthogonal eigenvectors $\hat{\bf{e}}_i$ with $\hat{\bf{p}}$. As shown in figure \ref{fig:pei}, both simulation and experiment suggest that the rod tends to be aligned with $\hat{\bf{e}}_1$ and $\hat{\bf{e}}_2$ and perpendicular to $\hat{\bf{e}}_3$, indicating that the rod primarily lies in the plane formed by $\hat{\bf{e}}_1$ and $\hat{\bf{e}}_2$.  

Another way to assess the relative contribution to the rod tumbling rate of vortical motion and strain is by conditioning on the magnitude of $\Omega_{ij}$ and $S_{ij}$, which can be represented by the enstrophy $\omega^2=\omega_i\omega_i$ and the dissipation rate $\epsilon=2\nu \langle S_{ij}S_{ij} \rangle$. Figure \ref{fig:rotrweps} shows the mean-squared rod tumbling rate conditioned on these two quantities. The tumbling rate monotonically increases by nearly two orders of magnitude with a similar dependence on both enstrophy and dissipation rate, indicating that the vorticity and the strain equally contribute to the tumbling rate of the rod on average. This two decade change of the conditional rod tumbling rate is larger than other dimensions, which also suggests that the enstrophy and the dissipation rate are the most important two dimensions.

In addition to the effect of the strain magnitude, we also investigated the effect of the orientation between the rod and the eigenframe of the strain-rate tensor on the tumbling rate, as shown in figure \ref{fig:condpei}. Recall that, in turbulence, the vorticity tends to be aligned with $\hat{\bf{e}}_2$ \citep{1987POFAshurst}. It would then be expected that the conditional average of the tumbling rate on $\langle\hat{\bf{p}}\cdot \hat{\bf{e}}_2\rangle$ (figure \ref{fig:condpei}(b)) would share some similarity with the conditional average on $\langle\hat{\bf{p}}\cdot \hat{\bf{\omega}}\rangle$ (figure \ref{fig:condpw}). In both cases, the tumbling rate monotonically decreases with increasing alignment, but the trend is much steeper for the case of $\langle\hat{\bf{p}}\cdot \hat{\bf{\omega}}\rangle$, because the vorticity vector plays a stronger role in determining the rod tumbling. The conditional average of the tumbling rate on $\langle\hat{\bf{p}}\cdot \hat{\bf{e}}_1\rangle$ (figure \ref{fig:condpei}(a)) is more complicated to explain. If the rod were perfectly parallel or perpendicular to $\hat{\bf{e}}_1$, the strain would act to stretch or compress the rod rather than to rotate it, and the tumbling rate due to strain would be zero in both limits. But if a rod were oriented at 45$^{\circ}$ with respect to $\hat{\bf{e}}_1$, the strain contribution to its tumbling rate would be maximized. So the squared tumbling rate should be small for both limits of alignment at $\langle\hat{\bf{p}}\cdot \hat{\bf{e}}_1\rangle=0$ and $\langle\hat{\bf{p}}\cdot \hat{\bf{e}}_1\rangle1$, but should have a peak near $\langle\hat{\bf{p}}\cdot \hat{\bf{e}}_1\rangle=cos(45^{\circ})=0.7$. Our results in figure \ref{fig:condpei}(a) are consistent with this expectation. The same argument can be made for the conditional average on $\langle\hat{\bf{p}}\cdot \hat{\bf{e}}_3\rangle$, as shown in figure \ref{fig:condpei}(c). In this case, however, the peak is much higher and the data are skewed towards $\langle\hat{\bf{p}}\cdot \hat{\bf{e}}_3\rangle=1$, as compared with the data conditioned on $\langle\hat{\bf{p}}\cdot \hat{\bf{e}}_1\rangle$.   Two factors are important here.  First, rods are rarely aligned with $\hat{\bf{e}}_3$, as seen in Fig.~\ref{fig:pei}.  Events where this occurs are likely to be events with large vorticity, so the rotation rate will be large.  In addition, the vorticity is preferentially perpendicular to $\hat{\bf{e}}_3$, and so a rod that is better aligned with $\hat{\bf{e}}_3$ gains a larger contribution to its tumbling from vortical motion. Both of these effects move the peak up and towards $\langle\hat{\bf{p}}\cdot \hat{\bf{e}}_3\rangle=1$. 

We have now described the dependence of rod tumbling rate on five independent dimensions: the magnitudes of the strain and the vorticity, captured by the dissipation rate and the enstrophy, respectively, and three independent characterizations of the rod orientation with respect to the velocity gradient tensor. All of them are important in determining the rod tumbling rate, and none can be neglected. The additional two dimensions necessary for deterring the problem concern the orientation of vorticity in the strain eigenframe, and can be taken to be, for example, $\langle{\hat{\bf{\omega}}}\cdot{\hat{\bf{e}}}_1\rangle$ and $\langle{\hat{\bf{\omega}}}\cdot{\hat{\bf{e}}}_2\rangle$. We find that the conditional mean square tumbling rate has only a very weak dependence on these two dimensions.  This may seem surprising since the relative orientation of vorticity and strain determine whether they reinforce or cancel each other and see in Fig.~\ref{fig:rot_rate}.  For prediction of the mean square tumbling rate, the parameter space may be able to be simplified to five dimensions, but it seems that a complete picture of the rotation of rods in turbulence will require specifying the full seven dimensional PDF of rod orientation and the velocity gradient tensor.

\section{Conclusion}

We have presented an experimental investigation of the tumbling rate of rods in turbulence. To assess the relative importance of various factors on the tumbling rate and to thereby estimate the effective dimensionality of the problem, we also simultaneously measured the full velocity gradient tensor near the rods. We obtained the gradient by implementing a scanning particle-tracking system, allowing us to image a high concentration of tracer particles. The quality of our velocity gradient measurements is comparable to previous experiments, and we were able to measure the trajectories of anisotropic particles at the same time.

We carefully explored the mean-squared tumbling rate conditioned on several different variables, including $\langle{\hat{\bf{p}}}\cdot {\bf{\hat{\bf{\omega}}}}\rangle$, $\epsilon$, $\omega^2$, $\langle\hat{\bf{p}}\cdot\hat{\bf{e}}_1\rangle$, $\langle\hat{\bf{p}}\cdot\hat{\bf{e}}_2\rangle$ and $\langle\hat{\bf{p}}\cdot\hat{\bf{e}}_3\rangle$. These variables were chosen to give a framework within which the relative contributions to the rod tumbling rate are easily interpretable. We found that the mean-squared tumbling rate was dependent on all five of these dimensions to some degree. As these variables can be separated into two classes, those that depend on the strain rate and those that depend on the vorticity, our results suggest that the rod tumbling rate depends approximately equally on $S_{ij}$ and $\Omega_{ij}$. 

We provide experimental evidence that rods are preferentially aligned with the vorticity, thus diminishing the potential contribution from $\Omega_{ij}$ and partially explaining why the measured rod tumbling rates are smaller than they would be if the rod orientation were random. We also found that the mean-squared rod tumbling rate monotonically decreases with increasing $\langle\hat{\bf{p}}\cdot\hat{\bf{e}}_2\rangle$, where $\hat{\bf{e}}_2$ is the intermediate strain-rate eigenvector, just as it does with $\langle\hat{\bf{p}}\cdot\hat{\bf{\omega}}\rangle$. The contribution from the local strain is small when the rod is either aligned with or orthogonal to $\hat{\bf{e}}_1$, but is largest when the rod is oriented at roughly $45^{\circ}$ with respect to $\hat{\bf{e}}_1$. The PDF of the rod tumbling rate conditioned on $\langle\hat{\bf{p}}\cdot\hat{\bf{e}}_3\rangle$ is similar to that conditioned on on $\langle\hat{\bf{p}}\cdot\hat{\bf{e}}_1\rangle$; the dependence is stronger, however, for $\langle\hat{\bf{p}}\cdot\hat{\bf{e}}_3\rangle$ because the rod feels the full effect of both the strain and the vorticity simultaneously when it is aligned with $\hat{\bf{e}}_3$. This result suggests that the intermittent tumbling rate of rods may be linked back to the geometric information contained in the relative orientation of a rod with the local velocity gradient. 

Finally, the new experimental technique we describe enables us to extract the flow motion near particles in a fully turbulent 3D system, and has the potential to be applied to many other problems such as larger particles where larger particle Reynolds number makes simulations much more difficult or the flow field generated by active anisotropic particles such as bacteria.

\begin{acknowledgements}
We thank Federico Toschi and Enrico Calzavarini for providing us with the DNS data.  We acknowledge support from U.S.~National Science Foundation grants DMR-1206399 and DMS-1211952 to Yale University and DMR-1208990 to Wesleyan University and COST Actions MP0806 and FP1005.
\end{acknowledgements}

\appendix
\section{Quality of experimental measurements}

\begin{figure}
\begin{center}
$\begin{array}{cc}
\includegraphics[width=5.4in]{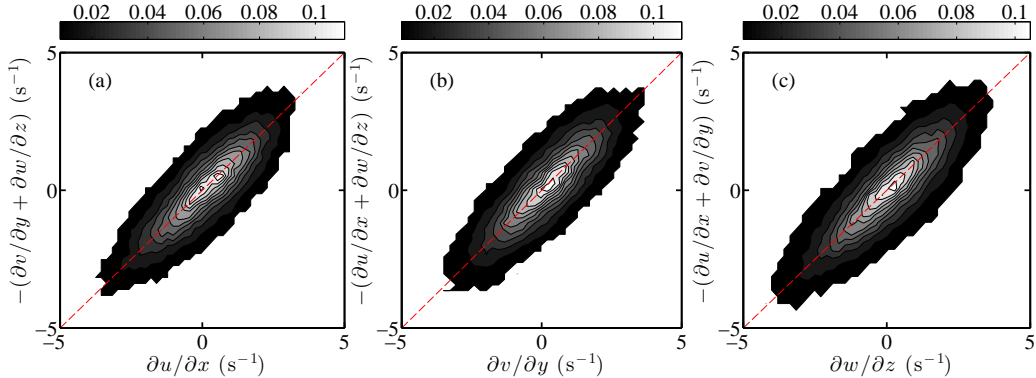}
\end{array}$
\caption{(Color online) Joint probability distribution function of one of the diagonal components of velocity gradient tensor versus the summation of the other two. The red dashed line represents the case for incompressible fluid, i.e. $\partial u_i/\partial x_i=0$.}
\label{fig:incomp}
\end{center}
\end{figure}

Here, we examine the quality of our experimental measurement of the velocity gradient tensor in several ways. First, we test the incompressible flow condition that requires the trace of the gradient to vanish.  Figure \ref{fig:incomp} shows the joint probability density function (jPDF) of one of the diagonal components of the velocity gradient tensor versus the sum of the other two. The shape of the contour lines for all panels look very similar, indicating that both the flow and the error are nearly isotropic. For perfectly incompressible flow, the vertical axes should be equivalent to the horizontal axes, as given by the red dashed line. Deviations from dashed line are caused by experimental uncertainties in measuring the velocity gradient.    These deviations can be quantified using the ratio between major and minor axes of the contour ellipses in the jPDF. The larger this ratio is, the more accurate the measurement of the velocity gradient tensor. For all three panels in figure \ref{fig:incomp}, this ratio is roughly 3.3, which is larger than the value of 2.6 in the early experiments of \citet{2005JFMLuthi}.  More recent experiments using scanning particle tracking systems by \citet{2005EIFHoyer} and \citet{2014MSTKrug} have measured deviations from incompressibility that are comparable or slightly smaller than ours.

\begin{figure}
\begin{center}
$\begin{array}{cc}
\includegraphics[width=5.4in]{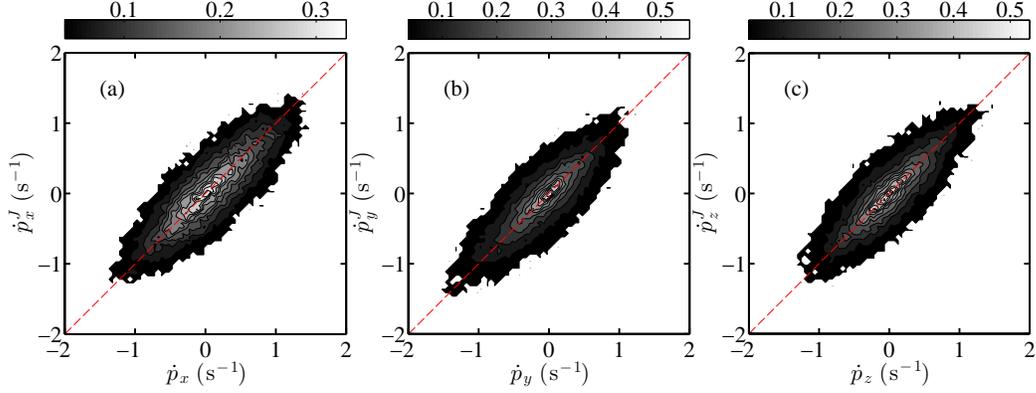}
\end{array}$
\caption{(Color online) Joint probability distribution function of the tumbling rate $\dot{p}_i$ from rod orientation measurements versus the tumbling rate calculated from Jeffery's equation $\dot{p}^J_i$ applied to the measured velocity gradient tensor. The red dashed line represents the case when two tumbling rates agree with each other, i.e. $\dot{p}_i=\dot{p}^J_i$}
\label{fig:compJeff}
\end{center}
\end{figure}

A more stringent and relevant test of our experimental accuracy is comparing the directly measured rod tumbling rate $\dot{p}_i$ computed by differentiating the time-dependent rod orientation with that calculated from Jeffery's equation $\dot{p}^J_i$ using the measured velocity gradient tensor. Figure \ref{fig:compJeff} shows the jPDF of $\dot{p}^J_i$ and $\dot{p}_i$. The contours of the jPDF again are ellipsoidal that  deviates from the linear red dashed line, just as they did in figure \ref{fig:incomp}. The ratio between the major and minor axes of the contours is also close to 3.3, indicating that the uncertainty in the measurements of the velocity gradient tensor is not amplified through Jeffery's equation. 

\begin{figure}
\begin{center}
$\begin{array}{cc}
\includegraphics[width=3.4in]{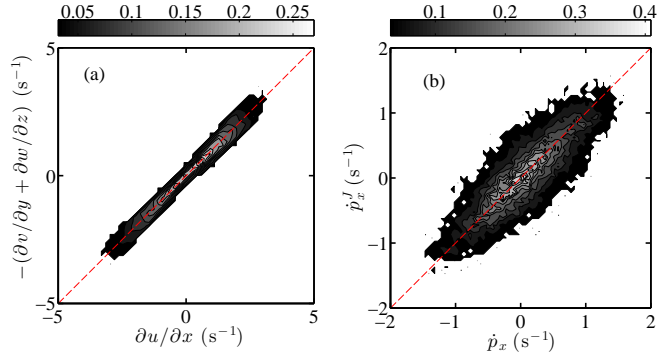}
\end{array}$
\caption{(Color online) Comparison of the joint probability distribution function with previous figures after setting the condition $\delta<0.15$ (a) same plot as figure \ref{fig:incomp} (a); (b) same plot as figure \ref{fig:compJeff} (a)}
\label{fig:omgcg}
\end{center}
\end{figure}

\begin{figure}
\begin{center}
$\begin{array}{cc}
\includegraphics[width=4.8in]{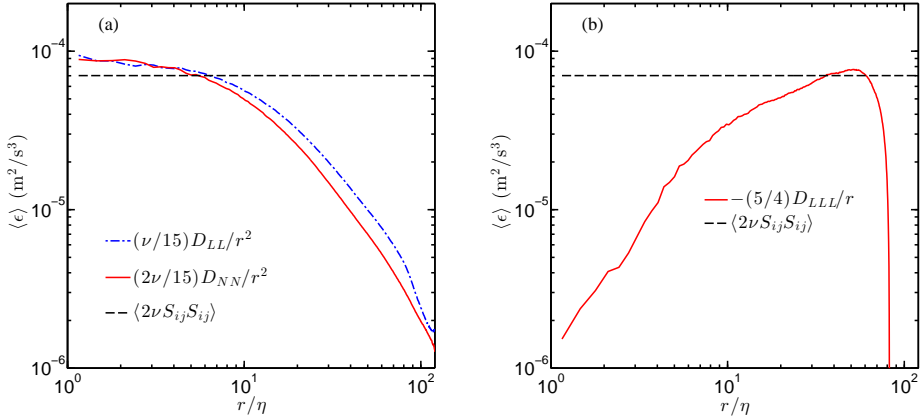}
\end{array}$
\caption{(Color online) The energy dissipation rate $\epsilon$ calculated from different methods (a) Comparison between $\epsilon$ obtained from the dissipative range of the longitudinal ($D_{LL}$, blue line) and transverse ($D_{NN}$, red line) second-order structure function and the direct measurement of velocity gradient tensor (black dashed line). (b) Dissipation rate obtained from compensated third-order structure function (solid line) and the direct measurement of the velocity gradient tensor (black dashed line).}
\label{fig:diss}
\end{center}
\end{figure}

Previously, it has been proposed that the incompressibility condition can be used as a weighting factor for smoothing trajectories \citep{2005JFMLuthi}. The apparent incompressibility is characterized by
\begin{equation}
\delta = \frac{|\frac{\partial u}{\partial x}+\frac{\partial v}{\partial y}+\frac{\partial w}{\partial z}|}{|\frac{\partial u}{\partial x}|+|\frac{\partial v}{\partial y}|+|\frac{\partial w}{\partial z}|}.
\end{equation}
The weighting factor is a function of this $\delta$; it is designed to be 1 for small $\delta$, and smoothly changes to 0 for large $\delta$. Thus if we smooth all the components of the velocity gradient using this weight, it nearly guarantees that the filtered velocity gradient have a small apparent compressibility. Note, however, that incompressibility only relies on the three diagonal components in velocity gradient tensor, which are independent of the other 6 components; thus, compressibility may not be directly related to the accuracy of full velocity gradient measurement. To test this argument, we plot in figure \ref{fig:omgcg}(a) the incompressibility test we used in figure \ref{fig:incomp} but only using samples with $\delta<0.15$. This sampling is equivalent to applying a weighting function that equals 1 for $\delta<0.15$ and 0 for $\delta>0.15$, similar to the one used before \citep{2005JFMLuthi}. After enforcing incompressibility, we would expect that the contours should lie closer to the linear dashed line. We also plot the jPDF of $\dot{p}^J_i$ and $\dot{p}_i$ in figure \ref{fig:omgcg}(b) as we did in figure \ref{fig:compJeff}, but again enforcing $\delta<0.15$. In this case, the shape of the contours does not change much, suggesting that weighting by the measured compressibility does not significantly improve the overall accuracy of the full velocity gradient tensor measurements.

\section{Dissipation rate measurement}

From the velocity gradient tensor, we can directly obtain the mean energy dissipation rate $\langle\epsilon\rangle=2\nu \langle S_{ij}S_{ij}\rangle$, one of the most important parameters in turbulence. For most experiments in homogeneous, isotropic turbulence without direct access to the velocity gradient tensor, the dissipation rate is extracted from the scaling of the second- or third-order Eulerian velocity structure functions in the inertial range. The velocity structure function of order $n$ is defined to be the $n^\mathrm{th}$ statistical moment of the velocity increment across a spatial scale $r$: $\langle[{\bf{u}}({\bf{x}}+{\bf{r}})-{\bf{u}}({\bf{x}})]^n\rangle$. Applying Kolmogorov's theory of isotropic turbulence \citep{K41}, the longitudinal $D_{LL}$ and transverse $D_{NN}$ components of the second-order structure function ($n=2$) scale as $D_{LL}=\langle\epsilon\rangle r^2/15\nu$ and $D_{NN}=2\langle\epsilon\rangle r^2/15\nu$ in the dissipative range $r<\eta$. Figure \ref{fig:diss}(a) shows both $D_{LL}$ and $D_{NN}$ compensated with $r^2/15\nu$ and $2r^2/15\nu$ respectively. There is a very short plateau near $r\approx \eta$, and the two compensated structure functions collapse with each other well in that range. Thus, the value of that plateau tells us the dissipation rate, which in our experiment is $\langle\epsilon\rangle\approx9\times10^{-5}$ m$^2/$s$^3$. In addition, figure \ref{fig:diss}(b) shows the third-order longitudinal structure function ($D_{LLL}$). The dissipation rate can be obtained from the Kolmogorov $4/5$ law ($D_{LLL}=-4\langle\epsilon\rangle r/5$) in the inertial range $\eta\ll r\ll L$, where $L$ is integral length scale. Here, because of the finite Reynolds number and the small measurement volume, there is a sharp cutoff near the infrared end of the inertial range. There is therefore only a narrow peak roughly from 50 $\eta$ to 60 $\eta$, but the estimate of $\langle\epsilon\rangle$ from that peak is very close to the estimate from the dissipation-range scaling of the second-order structure functions. The black dashed line in both panels shows the estimate from the velocity gradient measurement from tracer particles within a $\sim6\eta$ radius of the rod. It is clear that this value is slightly smaller than the estimate from both velocity structure function methods. This is because the velocity gradient is coarse-grained over a small volume, which removes contributions from some of the smallest scales to the energy dissipation rate. 

\bibliographystyle{jfm}

\end{document}